%% file: master_page.tex
\documentclass[useAMS,usenatbib]{mn2e}
\usepackage[a4paper,body={17cm,23cm},foot=2cm,head=15mm]{geometry}
\usepackage{graphicx}
\usepackage{times}
\usepackage{booktabs}
\usepackage{longtable}
\usepackage{amsmath}
\usepackage{amsfonts}
\usepackage{subfigure}
\usepackage{rotate}
\usepackage{url}
\usepackage{wasysym}
\usepackage{longtable}
\usepackage[title,titletoc]{appendix}
\usepackage{lscape}
\usepackage{rotating}
\usepackage{verbatim} 

\footnotesize
\newdimen\digitwidth    
\setbox0=\hbox{\rm0}
\digitwidth=\wd0
\catcode`!=\active
\def!{\kern\digitwidth}
\normalsize

\title[Pulsar searches in \textit{Fermi}-LAT unassociated sources with the Effelsberg
telescope]{Pulsar searches of \textit{Fermi} unassociated sources with the Effelsberg
telescope}

\author[E.~D.~Barr et al.]{\parbox{\textwidth}{
E.~D.~Barr,$^{1,2}$
L.~Guillemot,$^{1,3}$
D.~J.~Champion,$^{1}$
M.~Kramer,$^{4,1}$
R.~P.~Eatough,$^{1}$
K.~J.~Lee,$^{1}$
J.~P.~W.~Verbiest,$^{1}$
C.~G.~Bassa,$^{4}$
F.~Camilo,$^{5,6}$
\"O.~\c{C}elik,$^{7,8,9}$
I.~Cognard,$^{10}$
E.~C.~Ferrara,$^{7}$
P.~C.~C.~Freire,$^{1}$
G.~H.~Janssen,$^{4}$
S.~Johnston,$^{11}$
M.~Keith,$^{11}$
A.~G.~Lyne,$^{4}$
P.~F.~Michelson,$^{12}$
P.~M.~Saz~Parkinson,$^{13}$
S.~M.~Ransom,$^{14}$
P.~S.~Ray,$^{15}$
B.~W.~Stappers,$^{4}$
K.~S.~Wood$^{15}$}\vspace{0.4cm}
\\
\parbox{\textwidth}{$^{1}$ Max-Planck-Institut f\"ur Radioastronomie, Auf dem H\"ugel 69, 53121 Bonn, Germany\\
$^{2}$ email: ebarr@mpifr-bonn.mpg.de\\
$^{3}$ email: guillemo@mpifr-bonn.mpg.de\\
$^{4}$ Jodrell Bank Centre for Astrophysics, School of Physics and Astronomy, The University of Manchester, M13 9PL, UK\\
$^{5}$ Columbia Astrophysics Laboratory, Columbia University, New York, NY 10027, USA\\
$^{6}$ Arecibo Observatory, HC3 Box 53995, Arecibo, PR 00612, USA\\
$^{7}$ NASA Goddard Space Flight Center, Greenbelt, MD 20771, USA\\
$^{8}$ Center for Research and Exploration in Space Science and Technology (CRESST) and NASA Goddard Space Flight Center, Greenbelt, MD 20771, USA\\
$^{9}$ Department of Physics and Center for Space Sciences and Technology, University of Maryland Baltimore County, Baltimore, MD 21250, USA\\
$^{10}$ Laboratoire de Physique et Chimie de l'Environnement, LPCE UMR 6115 CNRS, F-45071 Orl\'eans Cedex 02, and Station de radioastronomie de Nan\c{c}ay, Observatoire de Paris, CNRS/INSU, F-18330 Nan\c{c}ay, France\\
$^{11}$ CSIRO Astronomy and Space Science, Australia Telescope National Facility, Epping NSW 1710, Australia\\
$^{12}$ W. W. Hansen Experimental Physics Laboratory, Kavli Institute for Particle Astrophysics and Cosmology, Department of Physics and SLAC National Accelerator Laboratory, Stanford University, Stanford, CA 94305, USA\\
$^{13}$ Santa Cruz Institute for Particle Physics, Department of Physics and Department of Astronomy and Astrophysics, University of California at Santa Cruz, Santa Cruz, CA 95064, USA\\
$^{14}$ National Radio Astronomy Observatory (NRAO), Charlottesville, VA 22903, USA\\
$^{15}$ Space Science Division, Naval Research Laboratory, Washington, DC 20375-5352, USA\\
}}

\date{Received: --Accepted:}

\begin{document}
\maketitle
\begin{abstract}

Using the 100-m Effelsberg radio telescope operating at 1.36 GHz, we
have performed a targeted radio pulsar survey of 289 unassociated
$\gamma$-ray sources discovered by the Large Area Telescope (LAT)
aboard the \emph{Fermi} satellite and published in the 1FGL catalogue
\citep{Abdo2010}.  This survey resulted in the discovery of
millisecond pulsar J1745$+$1017, which resides in a short-period binary system with a low-mass companion, $M_{c,min} \sim
0.0137 M_{\astrosun}$, indicative of ``Black Widow'' type systems. A
two-year timing campaign has produced a refined radio ephemeris,
accurate enough to allow for phase-folding of the LAT photons,
resulting in the detection of a dual-peaked $\gamma$-ray light-curve,
proving that PSR J1745$+$1017 is the source responsible for the
$\gamma$-ray emission seen in 1FGL~J1745.5$+$1018
\citep[2FGL~J1745.6$+$1015; ][]{Nolan2012}. We find the $\gamma$-ray
spectrum of PSR J1745$+$1017 to be well modelled by an
exponentially-cut-off power law with cut-off energy 3.2 GeV and photon
index 1.6. The observed sources are known to contain a further 10
newly discovered pulsars which were undetected in this survey. Our
radio observations of these sources are discussed and in all cases
limiting flux densities are calculated. The reasons behind the
seemingly low yield of discoveries are also discussed.
\end{abstract}


\input{Introduction}
\input{Source_selection}

\input{Observational_method}

\input{Sensitivity}

\input{Simulations}

\input{Results}

\input{Discussion}

\input{Conclusions}

\section*{Acknowledgements}
This work was carried out based on observations with the 100-m telescope of the
MPIfR (Max-Planck-Institut f\"{u}r Radioastronomie) at Effelsberg. 

The \emph{Fermi}-LAT Collaboration acknowledges generous ongoing support from a number
of agencies and institutes that have supported both the development and the
operation of the \emph{Fermi}-LAT as well as scientific data analysis. These include
the National Aeronautics and Space Administration and the Department of Energy
in the United States, the Commissariat \`{a} l'Energie Atomique and the Centre
National de la Recherche Scientifique/Institut National de Physique
Nucl\'{e}aire et de Physique des Particules in France, the Agenzia Spaziale
Italiana and the Istituto Nazionale di Fisica Nucleare in Italy, the Ministry of
Education, Culture, Sports, Science and Technology (MEXT), High Energy
Accelerator Research Organization (KEK), and Japan Aerospace Exploration Agency
(JAXA) in Japan, and the K. A. Wallenberg Foundation, the Swedish Research
Council, and the Swedish National Space Board in Sweden.

Additional support for science analysis during the operations phase is
gratefully acknowledged from the Istituto Nazionale di Astrofisica in Italy and
the Centre National d'\'{E}tudes Spatiales in France.

The Nan\c{c}ay Radio Observatory is operated by the Paris Observatory,
associated with the French Centre National de la Recherche
Scientifique(CNRS).

JPWV acknowledges support by the European Union under Marie-Curie
Intra-European Fellowship 236394. 

PCCF and JPWV acknowledge support by the European Research Council
under ERC Starting Grant Beacon (contract no. 279702).

We would like to thank Matthew Kerr for his input regarding initial source selection.

\bibliographystyle{mnras}
\bibliography{library.bib}
\newpage
\appendix
\input{PointingsAll2}

\end{document}

%% file: Introduction.tex
\section{Introduction}
\label{sec:intro}

The detection of pulsed $\gamma$-ray emission from the Crab Pulsar in
the early 1970's \citep{Vasseur1970,Grindlay1972}, the first of its
kind, brought new light to the study of pulsar emission physics and
high-energy emission physics in general. Gamma-ray photons of
energies greater than 100 keV are created in processes involving
nuclear or other non-thermal reactions, and as such become important
when exploring the Universe at its most energetic. The current model
for the creation of $\gamma$-ray photons that we see from pulsars
is that charged particles stripped from the surface are accelerated to
relativistic energies in the pulsar's strong electric field. As these
particles travel along the curved magnetic field lines, they produce
$\gamma$-ray photons via synchrotron radiation, curvature radiation
\citep[e.g.][]{Ruderman1975} and inverse Compton scattering from
lower-energy photons \citep[e.g.][]{Daugherty1986}. The study of these
processes gives insight into the structure and composition of the
magnetospheres of pulsars.

Prior to 2008, the most successful space-based $\gamma$-ray experiment was the Compton Gamma-Ray Observatory (CGRO), which was in orbit for nine years and carried the Energetic Gamma-Ray Experiment Telescope \citep[EGRET,][]{Kanbach1989}. EGRET was sensitive to $\gamma$-ray photons in the range 20 MeV -- 30 GeV, and during its lifetime brought the known number of $\gamma$-ray emitting pulsars up to at least six \citep{Thompson2008a}.  However, the legacy of EGRET for the radio community was not the pulsars it detected, but rather those sources for which it could make no positive association.  Targeted radio searches of these 169 $\gamma$-ray sources, unassociated with either pulsars or blazars, were performed, leading to several pulsar discoveries
\citep[e.g.][]{Keith2008,Champion2005}.

The Large Area Telescope (LAT) \citep{Atwood2009} aboard the \emph{Fermi} Gamma-ray Space Telescope, represents a significant improvement upon EGRET, providing a greater energy range and sensitivity, allowing for better measurements of source characteristics and localisations. With a host of new sources discovered, including many active galactic nuclei (AGNs) and pulsars, the \emph{Fermi} LAT is the most successful GeV $\gamma$-ray observatory to date. As with EGRET, it is those sources for which \emph{Fermi} cannot immediately provide an association that have piqued the interest of the pulsar searching community. A catalogue of 1451 $\gamma$-ray sources detected above 100 MeV was created from the first 11 months of LAT data. Of these sources, 630 were unassociated with known astrophysical objects \citep[AGNs, pulsars, etc.; ][]{Abdo2010}. Multi-wavelength observations of the unassociated sources were encouraged so as to determine their natures, with many radio observatories searching for radio pulsations in the \emph{Fermi} observational error ellipses \citep[e.g.][]{Keith2011,Ransom2011, Cognard2011}.

While \emph{Fermi} LAT data have already been proved to contain a wealth of pulsars, with more than 100 pulsars detected through blind periodicity searches and phase folding of LAT photons using known pulsar ephemerides \citep{Ray2011b}, low photon counts introduce strong selection biases in the detection of pulsars through blind searches of the LAT data. This is due to the
large amount of computation required to perform wide-parameter-space searches of sparse photon data sets. For this reason, blind searches of the LAT data currently have great difficulty in detecting millisecond pulsars (MSPs) or pulsars in binary systems.

Radio pulsation searches are subject to different biases and thus are an important alternate method for identifying LAT unassociated sources as pulsars. At the time of this writing, there have been 47 radio-loud pulsars discovered
through searches of these sources, of which 41 are MSPs likely to be
associated with their corresponding LAT source \citep{Ray2012}. These
discoveries highlight the importance of targeted radio searches of the
LAT data, as these pulsars were most likely undetected in more general
surveys due to shorter integration times or lack of searching for
binary motion.

Of the MSPs discovered, 10 are thought to be in `Black Widow'
systems where the companion star has a very low mass due to ablation
by the strong wind of the pulsar \citep{Fruchter1988}.  Before
\emph{Fermi} only three of these systems were known to exist outside
globular clusters \citep{Fruchter1988, Stappers1996, Burgay2006a},
which stresses the importance of investigating this new population of
pulsars uncovered by the LAT.  Those MSPs discovered which are
not in Black Widow systems may also be of great use to current and
future pulsar timing arrays for gravitational wave detection
\citep{Foster1990}, which benefit from an even distribution of
precisely timed pulsars across the sky.

In this paper we present a targeted search of 289 unassociated
\emph{Fermi} sources using the 100-m Effelsberg telescope operating at
1.36 GHz. The search has resulted in the discovery of a 2.65-ms
pulsar, PSR J1745$+$1017, in a 17.5-hour binary orbit with a 0.016-M$_{\astrosun}$ companion. The positions of 10 pulsars found in
other targeted searches of unassociated LAT sources are contained
within the 289 sources observed. For these sources, we discuss
possible reasons for our non-detections and provide flux density
limits where applicable.

This paper is structured as follows. In Section \ref{sec:sources}, we discuss selection criteria for
sources to be observed. In Section \ref{sec:obsmethod} we discuss the
observational methods and data processing. In Section
\ref{sec:sensitivity} we discuss the survey sensitivity. In Section
\ref{sec:sim} we discuss simulations of the survey. In Section
\ref{sec:results} we discuss the results of the survey. In Section
\ref{sec:discussion} we discuss the source selection and detection rate. In Section \ref{sec:conclusions} we present our
conclusions.

%% file: Source_selection.tex
\section{\emph{Fermi} catalogue source selection}
\label{sec:sources}

All the sources searched for this paper were selected either from the
\emph{Fermi} LAT First Source catalogue (1FGL) \citep{Abdo2010} or
from an unpublished update to the 1FGL which covered the first 18
months of \emph{Fermi} observations.

The 1FGL contains 630 sources unassociated with any known
astrophysical object, in the 100 MeV to 100 GeV energy range. The
catalogue includes source localisations defined in terms of an elliptical
fit to the 95\% confidence level, power-law spectral fits, monthly
light curves and flux measurements in 5 energy bands for each
source. A full description of the 1FGL can be found in
\citet{Abdo2010}. To narrow down the number of potentially observable
sources, those below $-20^{\circ}$ declination, and those for which
a known association existed were ignored. For all sources, the pointing
position was chosen to coincide with the centre of the error ellipse. As better source localisations became available, pointing positions were altered such that they coincided with the center of the updated error ellipse.

The remaining sources were ranked using an application of the Gaussian-mixture model as outlined in \citet{Lee2012}. The best pulsar candidates exhibit significantly curved emission spectra and little $\gamma$-ray flux variability over time, in contrast with e.g. blazars and other AGNs \citep[see e.g. Figure 17 of][]{Nolan2012}.

The top 250 ranked candidates from the 1FGL catalogue were selected to be observed. A further 39 highly ranked sources from the 18-month update to the 1FGL were also selected to be observed, giving a total of 289 target sources.

%% file: Observational_method.tex
\section{Observational method and data processing}
\label{sec:obsmethod}

All search data presented in this paper were taken with the 100-m
Effelsberg radio telescope at a centre frequency of 1.36 GHz between
November 2009 and July 2010, using the central horn of the new
Effelsberg multi-beam receiver with 300-MHz bandwidth.

The 289 sources selected for this survey were observed in three
different observing campaigns. Initially all sources were observed
with between 10- and 16-minute integrations.  These integrations
allowed for a preliminary shallow sweep of all sources such that the
brightest pulsars, those likely to have been of the greatest use for
timing applications, could be found.  The second observing campaign
was comprised of 32-minute integrations on 78 sources, with special
focus given to covering as many of the highest ranked sources as time
allowed. Finally, in the third campaign, 70 highly ranked sources were
observed with between 60- and 76-minute integrations, with 32 of these
sources observed multiple times.  Observing in this manner reduced the
effects of scintillation and of man-made radio-frequency interference (RFI), as most sources were observed multiple times with different integration times at different observing epochs.

Data were recorded over 512 filterbank channels of width 585.3 kHz
with a sampling rate of 53 $\mu$s \footnote{An in-depth review of the
  observational set-up for pulsar searches with the Effelsberg
  telescope will be published in the initial paper of the High Time
  Resolution Universe North pulsar survey (Barr et al., in
  prep.).}. Initially all data were sampled at 32 bits by the digitisers
before being brought down to 8 bits and written to high-capacity
magnetic tape for transportation and storage.

To meet the processing demands imposed by the 5.5 TB of data created
in the survey, a 22-node 168-core computing cluster situated in the
MPIfR was used for all data analysis. The \textsc{presto} software
package \citep{Ransom2001} was used for data processing.

In the first stage of the processing pipeline, data underwent RFI treatment in which a time- and
frequency-dependent mask was created to be applied at a later
stage. As the 1FGL catalogue contains no distance information, each beam
was de-dispersed at 2760 trial dispersion measures in the range 0--997
pc cm$^{-3}$ to mitigate against the frequency-dependent delay in the
pulsar signal due to dispersion by free electrons along the line of
sight. Our choice of such a fine sampling in dispersion space allows
for the retention of the data's maximum possible time resolution at
all dispersion measures. The effect of this was a heightened
sensitivity to millisecond and potential sub-millisecond pulsars.

All de-dispersed time series were fast-Fourier-transformed (FFT) and the
resulting power spectra were de-reddened and
known RFI frequencies were removed. To reconstruct power distributed through
harmonics in the Fourier domain, the process of incoherent harmonic summing was
used. Here the original spectra are summed with versions of themselves that have
been stretched by a factor of two such that all second-order harmonics are added
to their corresponding fundamental. This process was repeated four times such
that all power distributed in even harmonics up to the $16^{\rm{th}}$ harmonic may be
incoherently added to the fundamental \citep[see e.g.][]{handbook}. The spectra
from each stage of the summing process were searched for accelerated and
non-accelerated signals.

At this stage the number and size of the FFTs required to achieve
sensitivity to fast binaries becomes computationally too expensive for the
longest pointings. To deal with this, the processing of the data was split into
two stages. Initially all data were analysed at full length with a moderate
acceleration search in the Fourier domain. This analysis is very sensitive to
isolated and mildly accelerated pulsars in the data. The second processing stage
involves splitting the data into 10-minute blocks and re-analysing with a much
more intensive acceleration search. Although this stage uses shorter
integrations, it is more sensitive to highly accelerated binary systems in the
data. The details of the acceleration search can be found in Section 4.

Upon completion of the processing, all candidate signals underwent a
sifting routine that removes any signals which are likely to be RFI. 
Finally for the top 50 candidates a set of diagnostic plots were created for
visual inspection \citep[e.g.][]{Eatough2010}. In cases where there were more
than 50 candidates with greater than 8-$\sigma$ significance, the
pointing was considered to be contaminated by RFI and was flagged for
re-observation.

%% file: Sensitivity.tex
\section{Sensitivity}
\label{sec:sensitivity}
To estimate the sensitivity of our survey setup we used the radiometer equation
\citep[see e.g.][]{handbook},
\begin{equation} \label{eq:sensitivity}
S_{\rm{min}} = \beta \frac{S/N_{\rm{min}} T_{\rm{sys}}}{G \sqrt{n_p t_{\rm{obs}}
\delta f }} \left (
\frac{P_{\rm{cycle}}}{1-P_{\rm{cycle}}} \right )^{\frac{1}{2}} ,
\end{equation}
where the constant factor $\beta$ denotes signal degradation due to
digitisation, which for 8-bit digitisation is $\sim1\%$, giving
$\beta = 1.01$ \citep{Kouwenhoven2001}. $T_{\rm{sys}}$ is the system
temperature of the receiver. From flux density calibration
measurements we found $T_{\rm{sys}} = 25$ K. G is the antenna gain
(1.5 KJy$^{-1}$ at 1.36 GHz), $P_{\rm{cycle}}$ is the pulse duty
cycle, $t_{\rm{obs}}$ is the length of the observation, $\delta f$ is
the effective bandwidth of the receiver (240 MHz) and $n_p$ is the
number of polarisations summed, which for this survey is always 2. The
factor $S/N_{\rm{min}}$ is the minimum signal-to-noise ratio with
which we can make a detection. Based on false alarm statistics, we chose $S/N_{\rm{min}} = 8$.
\begin{figure}
\centering
\includegraphics[width=240pt,height=230pt]{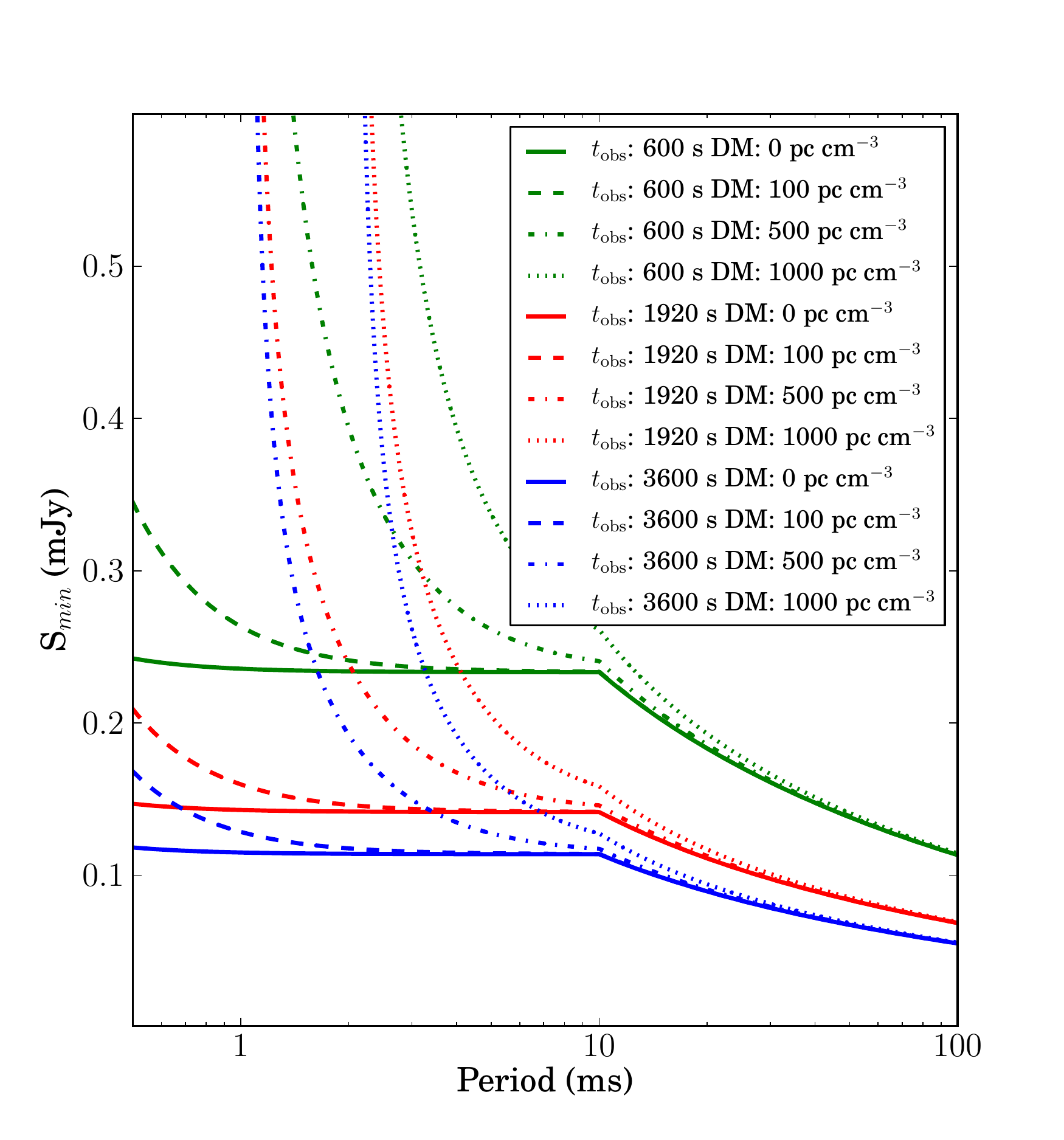}
\caption{Theoretical minimum detectable flux densities
  ($S_{\rm{min}}$) vs. spin period for three integration regimes. A
  minimum detectable signal-to-noise of 8 was imposed.
 The break point at period 10 ms occurs due to the assumption that
 the pulse duty-cycle scales as period$^{-0.5}$ with a maximum value
 of $1/3$ \citep{Kramer1998}.\label{fig:sensitivity}}
\end{figure}

Figure \ref{fig:sensitivity} shows the minimum detectable flux density
as a function of pulsar period for four dispersion measures; 0, 100,
500 and 1000 pc cm$^{-3}$. Assuming a pulsar of typical duty cycle
$5\%$ we achieve a minimum detectable flux density of 0.02 mJy for a
76-minute pointing and 0.06 mJy for a 10-minute pointing.

In a binary system, due to the Doppler effect, the apparent spin
frequency of the pulsar drifts with time, spreading the pulsar's power
in the Fourier domain. To reconstruct Fourier power smeared across
multiple bins, we employed \textsc{presto}'s \emph{accelsearch} routine which
uses the `correlation technique' of template matching in the Fourier
domain as outlined in \citet{Ransom2002}. The relationship between the
largest range of Fourier bins over which to search for drifted
signals, $Z_{\rm{max}}$, and the acceleration to which the search is
sensitive, $a_{0}$, can be described by
\begin{equation} \label{eq:accel}
 a_{0} = \frac{Z_{\rm{max}}Pc}{t^{2}_{\rm{obs}}} ,
\end{equation}
where $P$ is the spin period of the pulsar being searched for,
$t_{\rm{obs}}$ is the integration time and $c$ is the speed of light
in a vacuum.

By assuming a 1-ms period pulsar in a binary system we may determine
the sensitivity of the search at the extreme limit of the pulsar population. For the
first processing pass, where the data were analysed in full length, we
used $Z_{\rm{max}} = 50$ to achieve sensitivity to accelerations of up
to $|a_{0}| \sim 42$ m s$^{-2}$ and $|a_{0}| \sim 1$ m s$^{-2}$ for 10-
and 76-minute integrations respectively. For the second processing
pass, where data were analysed in 10-minute blocks, we used
$Z_{\rm{max}} = 600$ to achieve sensitivity to accelerations of up to
$|a_{0}| \sim 500$ m s$^{-2}$. For all systems which fall inside the
acceleration limits for their integration lengths, the minimum
detectable flux density may be calculated via Equation
\ref{eq:sensitivity}.

%% file: Simulations.tex
\section{Simulations}
\label{sec:sim}
As with any survey that contains many pointings, we may assume that the
chance detection probability for pulsars unassociated with the LAT sources
targeted is non-negligible. In order to determine this probability,
simulations of the normal pulsar and MSP populations were made based
on the model presented in \citet{Lorimer2006} using the
\textsc{psrpop} \footnote{http://psrpop.phys.wvu.edu/index.php}
software. As the population distributions for MSPs and normal pulsars
differ, separate simulations were performed for pulsars with rotational
periods above and below 40 ms.

To simulate the normal pulsar distribution, input model parameters were
chosen as follows:

\begin{itemize}
\item An empirical period distribution taken from the probability density function of the known
  population.
\item A log-normal luminosity distribution, with mean and variance in
  log space of $-$1.1 and 0.9, respectively \citep{FaucherGiguere2006}.
\item A Gaussian distribution of spectral indices, with mean of $-$1.6
  and variance of 0.5.
\item An exponential distribution for the height above the Galactic
  plane, with a scale height of 0.33 kpc \citep{Lorimer2006}.
\item A radial distribution as described in \citet{Lorimer2006}.
\item The NE2001 Galactic free electron density model
  \citep{Cordes2002}.
\end{itemize}

To simulate the MSP population, the same model parameters were used
with the Galactic scale height increased to 0.5 kpc to better match
the known MSP distribution \citep{Lorimer2006}.

The number of pulsars simulated was such that the cumulative number of pulsars
discovered in simulated versions of the Parkes Multibeam Pulsar Survey
\citep{Manchester2001}, the Swinburne Intermediate Latitude Pulsar Survey
\citep{Edwards2001} and its extension \citep{Jacoby2009}, and
the Parkes High Latitude Survey \citep{Burgay2006a}, matched the real
discovery numbers.

Each observation of the 289 sources in the survey was then compared against the
simulated pulsar distribution to determine if a chance detection could have been
made. The Galactic pulsar population was simulated 1000 times and comparisons
repeated with each simulation. We find a mean detection rate of 0.4 normal
pulsars and 0.04 MSPs in the 289 pointings. While not ruling out the possibility
of a chance MSP detection, these figures suggest that any MSP found in this
survey will be associated to the LAT source targeted at 96$\%$
confidence.

%% file: Results.tex
\section{Results}
\label{sec:results}
\subsection{PSR J1745$+$1017}

\subsubsection{Radio analysis} 

The main result of this work is the discovery of the radio pulsar PSR
J1745$+$1017 in LAT source 1FGL~J1745.5$+$1018
(2FGL~J1745.6$+$1015). Initially discovered as a mildly accelerated
candidate in a 10-minute data segment, PSR~J1745$+$1017 has a period
of 2.65 ms and a period derivative of $2.72\times 10^{-21}$, making it
the first MSP discovered with the Effelsberg telescope.

Upon discovery, PSR J1745$+$1017 was the subject of an intensive
timing campaign involving the Effelsberg, Lovell and Nan\c{c}ay radio
telescopes. Pulse times of arrival (TOAs) were analysed with the
\textsc{tempo2} software package \citep{Hobbs2006} to create a
phase-connected timing solution for the first two years of radio
observations. The pulsar ephemeris and a selection of derived
properties are displayed in Table \ref{tab:eph}.

From period and period derivative measurements we infer a
characteristic age of 16.9 Gyr. Combined with the rapid spin period of
PSR J1745$+$1017, this implies that the pulsar has undergone an
extensive period of accretion-induced spin-up due to mass transfer
from a companion star.  This hypothesis is supported by the timing
solution, which shows PSR J1745$+$1017 to be in a low-eccentricity
binary system with an orbital period of 17.5 hr and a low-mass
companion. Using the definition for the binary mass function,
\begin{equation} \label{eq:massfunc}
f(m_{1},m_{2}) = \frac{4\pi^{2}}{G}\frac{(a\sin{i})^{3}}{P_b^{2}} =
\frac{(m_{2}\sin{i})^{3}}{(m_{1}+m_{2})^{2}},
\end{equation}
where $m_{1}$ and $m_{2}$ are the masses of the pulsar and companion
respectively; $G$ is Newton's gravitational constant; $a \sin{i}$ is the
projected semi-major axis of the system, where $i$ is
the inclination angle of the binary (defined such that $i=90^{\circ}$
is edge on); and $P_{b}$ is the orbital period, we obtain $f(m_{1},m_{2}) = 1.4\times10^{-6}
M_{\astrosun}$. Assuming an average pulsar mass of 1.35
$M_{\astrosun}$, this gives minimum and median companion masses of 0.014
and 0.016 $M_{\astrosun}$ respectively.

This small range of low companion masses strongly suggests that PSR
J1745$+$1017 is a Black Widow system with a white dwarf companion
that has been heavily ablated by the strong particle wind from the
pulsar. Unlike Black Widow systems such as PSR B1957$+$21
\citep{Fruchter1988} and PSR J2051$-$0827 \citep{Stappers1996}, PSR
J1745$+$1017 exhibits neither eclipsing behaviour nor DM variations
across orbital phase, implying a low inclination angle. If the inclination is low, then the companion mass is likely to be larger than the presented value. As noted by \citet{handbook}, the probability of observing a binary system
with an inclination less than $i_0$ for a random distribution of
orbital inclinations is $p(i) = 1-\cos(i_0)$. This relation suggests that the inclination angle is greater than $26^{\circ}$ at 90\% confidence. We therefore place an upper limit of 0.032 $M_{\astrosun}$ on the mass of
the companion to the same confidence level.

From proper motion measurements of PSR J1754+1017, we find a
transverse velocity of $\sim 48\pm9$ km s$^{-1}$. The small
dispersion-measure-inferred distance to this pulsar, $d \sim 1.3$ kpc,
suggests that the Shklovskii effect \citep{Shklovskii1970}
contribution to the measured period derivative will be
non-negligible. For a pulsar with transverse velocity $v_{t}$, the
Shklovskii effect acts to increase the intrinsic period derivative of
the pulsar by a factor $Pv_{t}/cd $ \citep[see][]{Camilo1994}, where
$c$ is the speed of light in a vacuum. We find a contribution to the
period derivative of $5\pm2 \times 10^{-23}$ or about 18$\%$ of the
measured value.

Assuming a moment of inertia of $10^{45} $ g cm$^{2}$ and using the Shklovskii-corrected period derivative, we derive 
the spin-down luminosity of PSR J1745$+$1017 to be $\dot{E} =
4\pi^{2}I\dot{P}/P^{3} = 4.7 \times 10^{33} $ ergs s$^{-1}$, a value
which makes this pulsar a good candidate for pulsed $\gamma$-ray emission
\citep{Abdo2010}.

\begin{table}
\centering
\caption{PSR J1745+1017 ephemeris created from TOAs taken with the Nan\c{c}ay,
Effelsberg and Lovell telescopes over the course of 22 months. Numbers in
parentheses represent twice the formal 1-$\sigma$ uncertainties in the
trailing digit as determined by \textsc{tempo2}. The dispersion-measure-derived distance was estimated using the NE2001 Galactic electron density model \citep{Cordes2002}. Due to intrinsic uncertainties in this model, the estimation is likely to have an uncertainty of $\sim 20\%$. The mass function calculation assumes an
average mass of 1.35 $M_{\astrosun}$ for the pulsar. The characteristic age, spin-down luminosity and surface magnetic field strengths were calculated using the Shklovskii-corrected period derivative. The position, frequency and DM are all measured with respect to the given reference epoch. These parameters were
determined with \textsc{tempo2}, which uses the International Celestial
Reference System and Barycentric Coordinate Time. Refer to \citet{Hobbs2006} for
information on modifying this timing model for observing systems that use
\textsc{tempo} format parameters.\label{tab:eph}}

\begin{tabular}{lr}
\hline
PSR J1745+1017 ephemeris\\
\hline
\hline
\\
Fitted timing parameters\\
\hline
Right Ascension (R.A. J2000) (hh:mm:ss) & 17:45:33.8371(7)\\
Declination (Decl. J2000) ($\,^{\circ}:':''$) & +10:17:52.523(2)\\
Proper motion:&\\
in R.A. ($\mu_{\alpha} \cos(\rm{Decl.})$) (mas yr$^{-1}$) & 6(1)\\
in Decl. ($\mu_{\delta}$) (mas yr$^{-1}$) & -5(1) \\
Period (s) & 0.00265212967108(3)\\
Period derivative ($\times 10^{-21}$) & 2.73(1)\\ 
Dispersion measure (pc cm$^{-3}$) &  23.970(2)\\
Orbital period (days) & 0.730241444(1)\\
Projected semi-major axis (lt-s) & 0.088172(1)\\
Epoch of ascending node (MJD) & 55209.968794(2)\\
$\kappa$ ($\equiv e\cos{\omega}$) ($\times 10^{-5}$) & 0(2)\\
$\eta$ ($\equiv e\sin{\omega}$) ($\times 10^{-5}$) & 0(2)\\
\\
Fixed parameters\\
\hline
Reference epoch (MJD) & 55400\\
Clock correction procedure & TT(TAI)\\
Time system & TCB\\
Solar system ephemeris model & DE414\\
Binary model & ELL1 \citep{Lange2001}\\
\\
Derived parameters\\
\hline
Frequency (Hz) & 377.05547013813(5)\\
Frequency derivative (Hz s$^{-1} \times 10^{-16}$) & -3.88(2)\\
Orbital eccentricity ($\times 10^{-5}$) & 0(2)\\
Epoch of periastron passage (MJD) & 55210.3(4)\\
Galactic longitude (J2000) (deg) & 34.8693081(3)\\
Galactic latitude (J2000) (deg) & 19.2536887(5)\\
Mass function (M$_{\astrosun}$) & 1.38(1)$\times10^{-6}$\\
Minimum companion mass (M$_{\astrosun}$) & 0.0137\\
Median companion mass (M$_{\astrosun}$) & 0.0158\\
Dispersion measure-derived distance (kpc) & 1.3(2)\\
Shklovskii-corrected & \\
period derivative ($\times 10^{-21}$) & 2.22(5)\\
Characteristic age (Gyr) & 18.9\\
Spin-down luminosity ($\times 10^{33}$ erg s$^{-1}$) & 4.7\\
Surface magnetic field ($\times 10^{7}$ G) & 7.7\\
rms residual ($\mu s$) & 5.05\\
\\
Further parameters\\
\hline
Median flux density at 1.36 GHz (mJy) & 0.3\\
Maximum flux density at 1.36 GHz (mJy) & 4.4\\
Span of timing data (MJD) & 55225 - 56026\\
Number of TOAs & 156\\
\hline
\end{tabular}
\end{table}

\input{GammaAnalysis}

\subsection{Radio pulsar non-detections}

During the course of the observations, other observatories discovered six radio-emitting pulsars associated with the sources observed in this survey. These pulsars were not detected in this survey due to
a combination of reasons, such as limited flux density, local RFI
conditions and source position redefinitions. Further information on
all sources observed in this work may be found in the corresponding
on-line material, with the 15 highest ranked sources presented in Table
\ref{tab:alldata}.

\subsubsection{PSR J2030+3641}

PSR~J2030+3641 \citep{Camilo2012}, initially undetected in our data
processing, exhibits a rotational frequency which is similar to the
10$^{\rm{th}}$ sub-harmonic of the 50-Hz local mains frequency. Unfortunately
signals which are similar in frequency to harmonics of the mains are
often lost in noise or excised from the data in the process of Fourier
RFI excision, wherein known RFI signals are removed from the power
spectrum of the observation prior to pulse detection. It should be
noted that this pulsar was discovered using the Robert C. Byrd Green
Bank Telescope in the US, where the mains frequency is 60 Hz.

Reprocessing of these data with the published ephemeris yielded a
detection of the pulsar at signal-to-noise 6.4, a value which lies
below the detection threshold of this survey. Using Equation
\ref{eq:sensitivity}, we found a corresponding estimated flux density
of 0.08 mJy. This value is half that of the published flux density at
1.5 GHz, a discrepancy which may be attributed to the extra noise
induced by the oscillations of the mains electricity at the folding
period. It should be noted, that the derived position of J2030+3641 is
coincident with the position observed in this survey.

\subsubsection{PSRs J1646$-$2142, J1816$+$4510 and J1858$-$2218}

PSRs J1646$-$2142, J1858$-$2218 \citep{Ray2012} and J1816$+$4510
\citep{Kaplan2012} were all discovered at LAT positions that were
significantly different from the position determined in the
1FGL catalogue. Because of this, the discovery position of
these pulsars is a half-beamwidth or more away from the position
observed in this work, and so no detection is expected.

\subsubsection{PSRs J0307+7443 and J1828+0625}
\label{missed-radio} 
PSRs J0307$+$7443 and J1828$+$0625 \citep{Ray2012} were both
discovered at positions coincident with sources observed in
this survey. Both pulsars were detected at low radio frequency and
appear to be weak radio emitters. Assuming similar profile
characteristics at both 1.36 GHz and at the discovery frequency, we
found upper limits of 0.1 and 0.2 mJy on the radio flux densities at 1.36 GHz
for PSRs J0307$+$7443 and J1828$+$0625, respectively.

\subsection{Gamma-ray pulsar non-detections}
Through blind searches of \emph{Fermi} LAT photons, a further four
pulsars associated with the sources observed have since been
discovered \citep{Pletsch2012,Parkinson2011}.  No radio detection has
so far been made for these pulsars, therefore in Table
\ref{tab:missed-gamma} we present our upper limits on their radio flux
densities at 1.36 GHz.

A simple model for the telescope beam was used to adjust
limiting flux densities for position offsets between the pointing and true
position of the source. Limiting flux densities were divided by a
factor $q$, given by $q = e^{-(\theta/\phi)^2/1.5}$, where $\theta$ is the
pointing offset and $\phi$ is the beam half-width at half
maximum, to return the true limiting flux density at 1.36 GHz (S$_{\rm{min}}$).

\begin{table}
\centering
\begin{tabular}{lccr}
\hline
PSR & Pointing & $S_{\rm{min}}$ & Reference\\
& offset ($^\circ$) & (mJy)& \\
\hline
\hline
 J0734$-$1559 & 0.05 & 0.1 & \citet{Parkinson2011} \\
 J1803$-$2149 & 0.05 & 0.1 & \citet{Pletsch2012} \\
 J2030+4415 & 0.02 & 0.08 & \citet{Pletsch2012} \\
 J2139+4716 & 0.04 & 0.09 & \citet{Pletsch2012} \\
\hline
\end{tabular}
\caption{Limiting flux densities ($S_{\rm{min}}$) for LAT detected pulsars
coincident
with sources observed in this work. All pulsars are assumed to have a radio
pulse with a 10$\%$ duty-cycle and scattering effects are ignored. It should be
noted that the upper radio flux limits presented here are higher than those
published in \citet{Pletsch2012}, due to the use of an increased detection
threshold of signal-to-noise 8. \label{tab:missed-gamma}}
\end{table}

%% file: GammaAnalysis.tex
\subsubsection{Gamma-ray analysis}
\label{gamma_ana}

In order to characterize the $\gamma$-ray emission of PSR~J1745+1017,
we selected \emph{Fermi} LAT data recorded between 2008 August 4 and
2012 March 7, with reconstructed energies larger than 0.1 GeV,
directions within a circular region of interest (ROI) of 15$^\circ$
radius around the pulsar's position, and zenith angles smaller than
100$^\circ$. We further restricted the dataset to ``Source'' class
events of the P7\_V6 instrument response functions, and rejected times
when the rocking angle of the LAT exceeded 52$^\circ$ or when the
Earth's limb infringed upon the ROI. The selected $\gamma$-ray events
were finally phase-folded using the ephemeris given in
Table \ref{tab:eph} and the \emph{Fermi} plug-in distributed
with \textsc{tempo2} \citep{Ray2011a}. 

Initial pulsation searches using standard data selection cuts yielded 
marginal detections only. For instance, selecting photons found within 
1$^\circ$ of the pulsar and with energies larger than 0.1 GeV, we found an 
\emph{H}-test parameter \citep{deJager2010} of 15.6, which translates 
to a significance of $\sim$ 3.1$\sigma$. Nonetheless, pulsation searches 
can be made more sensitive by weighting the photons by the probability 
that they originate from the pulsar. These probabilities can be computed 
through a spectral analysis of the pulsar and the neighbouring 
sources \citep{Kerr2011,Guillemot2012}.

The $\gamma$-ray spectrum of PSR~J1745+1017 was measured by fitting sources in
the ROI using a binned likelihood method, with the \emph{pyLikelihood} module
included in the \emph{Fermi} Science Tools\footnote{http://fermi.gsfc.nasa.gov/ssc/data/analysis/scitools/overview.html}.
The source model used for the analysis included the spectral
parameters of the 78 sources of the \emph{Fermi} LAT Second Source 
Catalogue \citep[2FGL][]{Nolan2012} found within $20^\circ$ of the pulsar. 
The spectrum of PSR~J1745+1017 was represented by an exponentially-cut-off power-law
of the form $dN/dE = N_0 \left( E / \mathrm{1\ GeV} \right)^{-\Gamma} \exp \left(- E / E_c \right)$, where $N_0$ is a normalization factor, $\Gamma$ is
the photon index, and $E_c$ is the cut-off energy. The extragalactic
diffuse emission and the residual instrument background were modelled
using the \emph{iso\_p7v6source} template, and the Galactic diffuse
emission was modelled using the
\emph{gal\_2yearp7v6\_v0} map cube. The spectral parameters of the 
sources within 5$^\circ$ of the pulsar as well as the normalizations
of the diffuse models were re-fit, while the parameters of other 
sources were fixed at the 2FGL catalogue values. 
We found no evidence of significant emission from the pulsar in the phase range [0.6; 1]. In order to increase the signal-to-noise ratio of the pulsar we thus restricted the dataset to photons with reconstructed pulse phases between 0 and 0.6 (the $\gamma$-ray light curve of PSR~J1745+1017, presented below, is indeed compatible with showing emission only in this phase range).
The best-fitting spectral parameters of PSR~J1745+1017 are listed
in Table \ref{gammaparams}, along with the integrated photon and
energy fluxes above 0.1 GeV derived from these results. Systematic 
uncertainties were calculated by running the same analysis as described 
above, but using bracketing IRFs for which the effective area has been 
perturbed by $\pm$10\% at 0.1 GeV, $\pm$5\% near 0.5 GeV, and $\pm$10\% at 10 GeV, 
with linear interpolations in log space between.

\begin{figure}
\centering
\includegraphics[width=0.95\columnwidth]{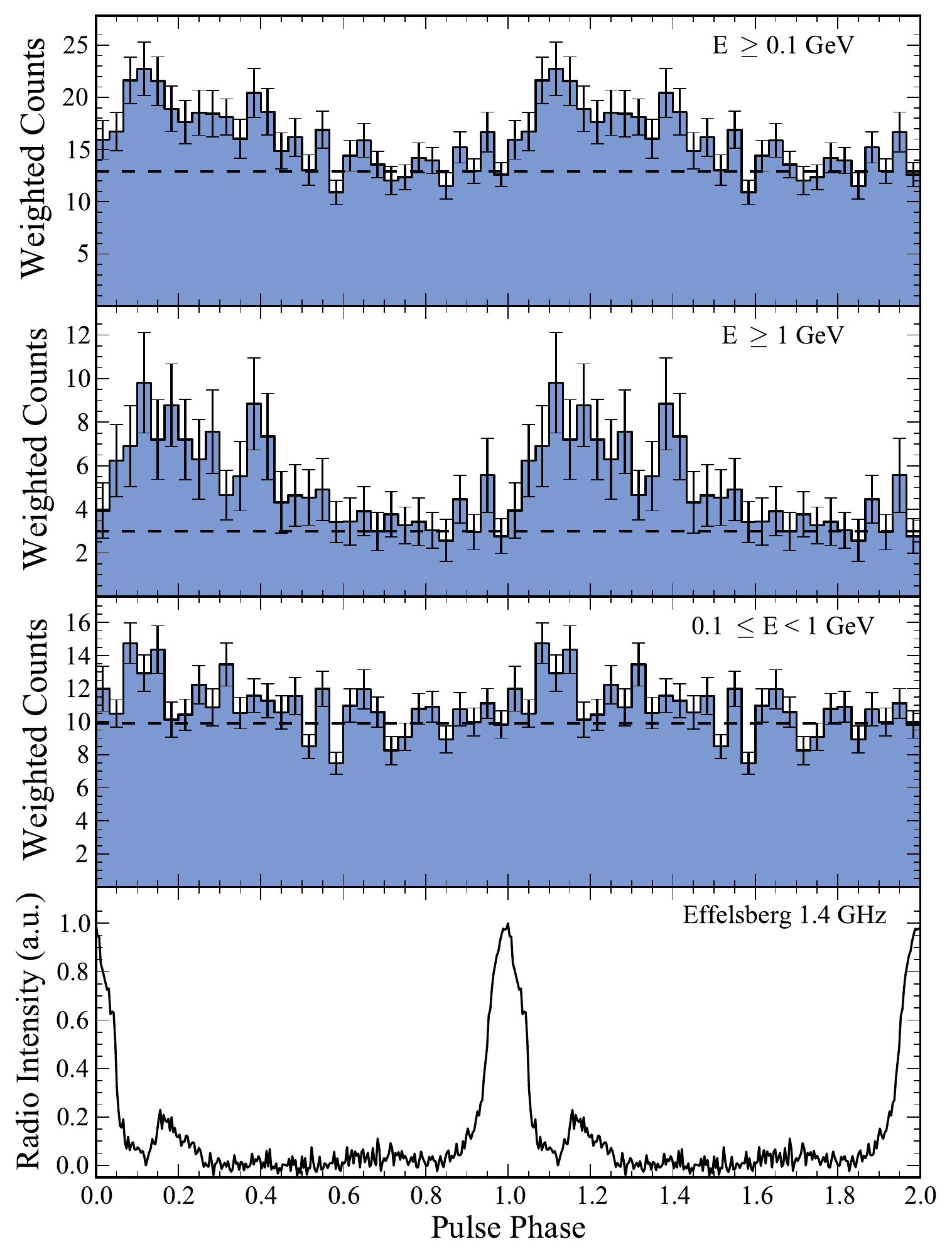}
\caption{Multi-wavelength light curves of PSR~J1745+1017. The bottom panel shows
the 1.4 GHz radio profile recorded with the Effelsberg Radio Telescope. The upper
panels show probability-weighted $\gamma$-ray light curves in different energy
ranges, with 30 bins per rotation. Horizontal dashed lines show the estimated
$\gamma$-ray background levels. Two rotations are shown for clarity.\label{lc1745gamma}}
\end{figure}

Using the best-fitting spectral model for the ROI obtained from the analysis
described above, we could calculate probabilities that the photons in the ROI
originate from PSR~J1745+1017. Selecting events found within $5^\circ$ and 
with calculated probabilities larger than 0.01, we
obtained a weighted $H$-test parameter \citep[see][]{Kerr2011} of 58.0,
corresponding to a pulsation significance of 6.5$\sigma$. Figure
\ref{lc1745gamma} shows probability-weighted light curves for PSR~J1745+1017 in
different energy bands. The background levels shown in Figure \ref{lc1745gamma}
were calculated by summing the probabilities that selected $\gamma$-ray events
do not originate from the pulsar, as described in \citet{Guillemot2012}.
Statistical error bars were obtained by calculating $\sqrt{\sum_i w_i^2}$, where
$w_i$ represents the photon probability and $i$ runs over photons falling in the
same phase bin \citep{Pletsch2012}. As can be seen from Figure
\ref{lc1745gamma}, the $\gamma$-ray profile of PSR~J1745+1017 shows evidence for
two distinct peaks, at phases $\sim 0.14$ and $\sim 0.39$.
Fits of the two $\gamma$-ray peaks with Lorentzian pulse shapes above 0.3 GeV yielded the peak positions $\Phi_i$
and the full widths at half maxima FWHM$_i$ listed in Table \ref{gammaparams}. We also attempted to fit the light curve with one asymmetric Lorentzian pulse shape, and found that the model with two peaks is slightly preferred, at the $\sim 1\sigma$ level.
The radio-to-$\gamma$-ray lag $\delta = \Phi_1 - \Phi_r$ (where $\Phi_r = 0$ 
is the phase of the maximum of the radio profile shown in Figure \ref{lc1745gamma}) 
and the $\gamma$-ray peaks separation $\Delta = \Phi_2 - \Phi_1$ are found to be 
$\delta = 0.14 \pm 0.04$ and $\Delta = 0.26 \pm 0.06$ (see Table \ref{gammaparams}). 
The uncertainty on the radio-to-$\gamma$-ray lag due to the error on the measurement 
of the DM parameter is estimated to be $\Delta(\delta) = \Delta(\mbox{DM}) / (K f^2)$ 
where $K = 2.410 \times 10^{-4}$ MHz$^{-2}$~cm$^{-3}$~pc~s$^{-1}$ is 
the dispersion constant. We find $\Delta(\delta) \sim 10^{-3} \times P$, which 
is very small compared to the statistical error bar. 
Such values of $\delta$ and $\Delta$ are relatively
common amongst other known $\gamma$-ray pulsars \citep[see Figure 4
of][]{Abdo2010a}, and match the predictions of theoretical models that place
the high-energy emission from pulsars at high altitudes in the
magnetosphere \citep{Romani1995}.

\begin{table}
\centering
\begin{tabular}{lc}
\hline
Parameter & Value \\
\hline 
\hline
First peak position, $\Phi_1$ \dotfill & $0.14 \pm 0.04$ \\
First peak full width at half-maximum, FWHM$_1$ \dotfill & $0.20 \pm 0.16$ \\
Second peak position, $\Phi_2$ \dotfill & $0.39 \pm 0.03$ \\
Second peak full width at half-maximum, FWHM$_2$ \dotfill & $0.10 \pm 0.10$ \\
Radio-to-$\gamma$-ray lag, $\delta$ \dotfill & $0.14 \pm 0.04$ \\
$\gamma$-ray peak separation, $\Delta$ \dotfill & $0.26 \pm 0.06$ \\
\hline
Photon index, $\Gamma$\dotfill & $1.6 \pm 0.2_{-0.1}^{+0.1}$ \\
Cutoff energy, E$_c$ (GeV)\dotfill & $3.2 \pm 1.2_{-0.1}^{+0.2}$ \\
Photon flux ($> 0.1$ GeV), $F_{100}$ (10$^{-8}$ cm$^{-2}$ s$^{-1}$)\dotfill & $1.1 \pm 0.3_{-0.1}^{+0.1}$ \\
Energy flux ($> 0.1$ GeV), $G_{100}$ (10$^{-12}$ erg cm$^{-2}$ s$^{-1}$)\dotfill & $9.3 \pm 1.2_{-0.6}^{+0.3}$ \\
Luminosity, L$_\gamma$ / f$_\Omega$ (10$^{33}$ erg s$^{-1}$)\dotfill & $1.8 \pm 0.6_{-0.6}^{+0.6}$ \\
Efficiency, $\eta$ / f$_\Omega$\dotfill & $0.3 \pm 0.1_{-0.1}^{+0.1}$ \\\hline
\end{tabular}
\caption{Measured $\gamma$-ray light curve and spectral parameters for PSR
J1745+1017. First quoted uncertainties are statistical, and the second are systematic. Details on the determination of the systematic uncertainties are given in Section~\ref{gamma_ana}.\label{gammaparams}}
\end{table}

\begin{figure}
\centering
\includegraphics[width=0.95\columnwidth]{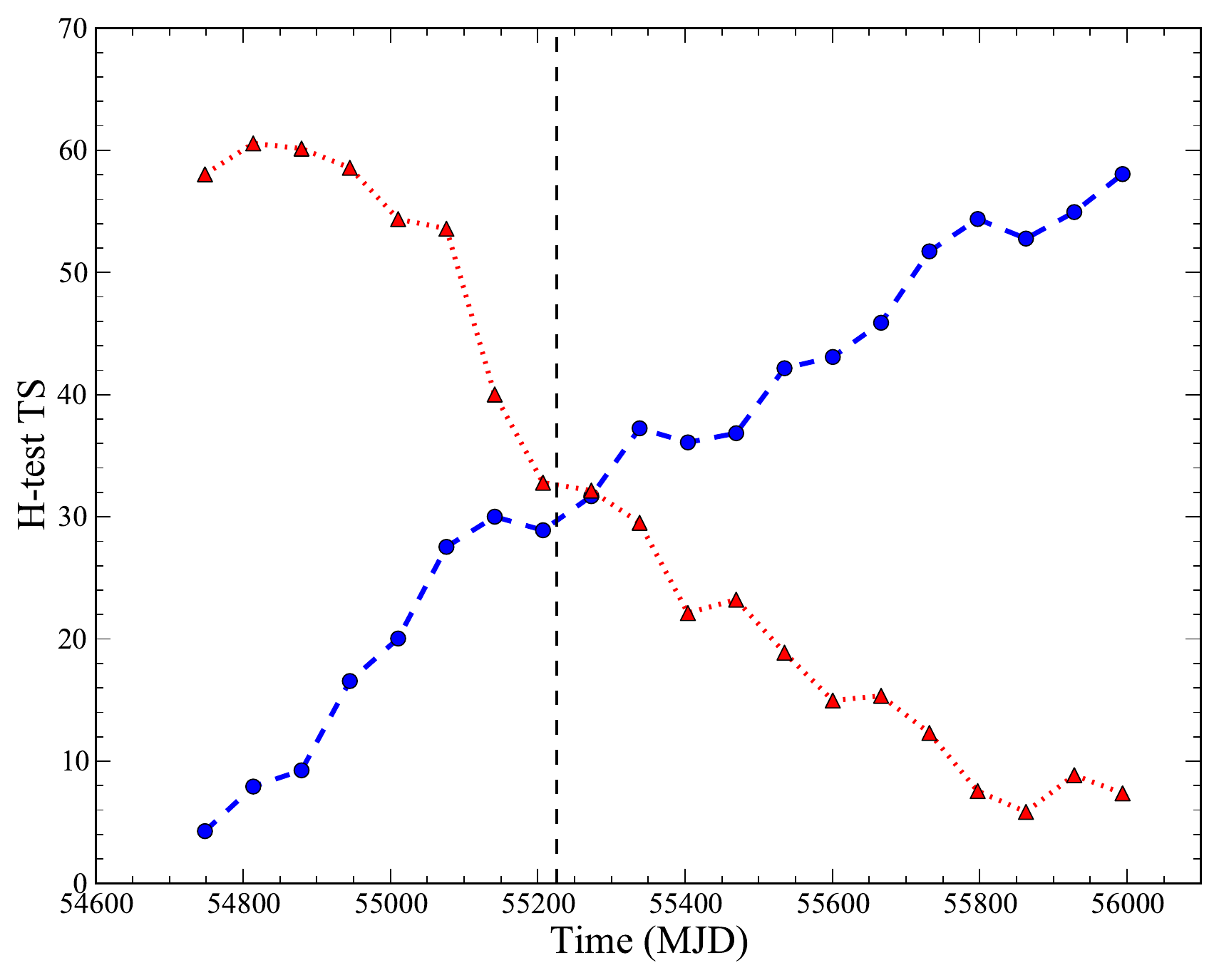}
\caption{Evolution of the weighted $H$-test Test Statistic (TS) as a function of
time. The blue, dashed line shows the weighted $H$-test parameter with increasing
time, using data taken from 2008 August 8. The red, dotted line shows the
weighted $H$-test parameter when going backwards in time, using data taken
before 2012 March 7. The vertical, dashed line indicates the formal start of
the ephemeris validity interval. \label{TS_time}}
\end{figure}

The ephemeris used for phase-folding the $\gamma$-ray data considered
in this analysis is based on radio timing taken after 2010 January 30 (MJD 55226). 
In an attempt to determine whether the ephemeris
describes the rotational behaviour of the pulsar over the
entire \emph{Fermi} LAT dataset accurately, we analysed the evolution
of the weighted $H$-test parameter as a function of time. The results
of this analysis are shown in Figure \ref{TS_time}. First of all,
because the $H$-test depends linearly on the number of photons for a
given pulsed signal fraction, the linear increase of the $H$-test
parameter as a function of the dataset length provides further
evidence that the pulsed $\gamma$-ray signal from PSR~J1745+1017 is
real. Moreover, the increase of the $H$-test parameter when going
forward or backward in time is monotonic outside the formal
ephemeris validity interval, which indicates that the ephemeris given
in Table \ref{tab:eph} provides a good description of the pulsar's 
rotational behaviour across the entire LAT dataset. 

From the energy flux $G_{100}$ measured from the spectral analysis, we
calculated the $\gamma$-ray luminosity above 0.1 GeV using $L_\gamma = 4 \pi
f_\Omega G_{100} d^2$, where $f_\Omega$ is a geometrical correction factor
depending on the beaming angle of the pulsar and the viewing geometry
\citep{Watters2009}. Using the distance derived from the NE2001 model of $d =
1.3 \pm 0.2$ kpc, we get the $\gamma$-ray luminosity 
$L_\gamma / f_\Omega \sim 1.8 \times 10^{33}$ erg s$^{-1}$ and the 
$\gamma$-ray efficiency, $\eta / f_\Omega = L_\gamma / \dot E \sim 0.3$.
This value for the efficiency lies well within the distribution of 
$\gamma$-ray efficiencies seen for other LAT-detected MSPs \citep{Abdo2010a}. 

From the measurement of the $\gamma$-ray energy flux and the proper motion of the pulsar, we can put an upper limit on the distance derived from the inequality $L_\gamma < \dot E$ \citep[see Equation 2 of ][]{Guillemot2012}. Assuming $I = 10^{45}$ g cm$^2$ and $f_\Omega = 1$ \citep[a typical value for other $\gamma$-ray MSPs, see e.g. ][]{Venter2009}, we get $d \leqslant 1.9$ kpc.

%% file: Discussion.tex
\section{Discussion}
\label{sec:discussion}

Considering the success that other surveys of unassociated \emph{Fermi} LAT
sources have had \citep{Keith2011,Cognard2011,Ransom2011,Camilo2012},
we must consider the reasons behind our relatively small yield of
discoveries.

Sources observed in this work were selected based on information
provided in the 1FGL catalogue. However, in the recently published
2FGL catalogue, many of these sources are no longer listed. A detailed description of the possible causes for this can be found in Section 4.2 of \citet{Nolan2012}.
Using the 2FGL catalogue to examine the 289 sources selected for
observation, we find:

\begin{itemize}
\item 106 sources that are no longer listed.
\item 46 sources that have moved by more than half the FWHM of the
  telescope beam.
\item 22 sources marked as potentially spurious.
\end{itemize}

Selecting against these sources and choosing only sources for which
the entire updated 95\% confidence region was contained within the
FWHM of the telescope beam, we are left with 72 sources for which
observations adequately cover the LAT source targeted.

This would suggest that the detection rate for the survey is
$\sim$1.4\%. However, as noted by \citet{Camilo2012}, a large number
of pulsar surveys at radio wavelengths have focused on observing the
Galactic plane at low latitudes, introducing a strong bias against the
discovery of new radio-emitting pulsars in this volume. Selecting only
those 58 sources that lie in the region $ |b| > 3.5^\circ$, we find
the detection rate to be ~1.7\%.

We note that the strategy employed by other surveys of LAT
unassociated sources has generally been to select a small number of
sources and observe with long integration times, often at lower
frequencies \citep[e.g.][]{Ransom2011}. By contrast, we have selected
a large number of sources and observed at various integration lengths
at higher frequency. Although this strategy has lacked the discoveries
that other searches have made, it has provided the most complete
picture yet of the LAT sky at high radio frequencies.

The results of our survey suggest that future surveys of unassociated LAT sources would benefit from lower observing frequencies, as wider beam-widths will increase the LAT error ellipse coverage, which, when combined with the decrease in LAT error ellipse size due to increasing observation depth, will significantly reduce the problems associated with source redefinitions. Furthermore, we note that the majority of pulsars discovered through radio searches of the LAT data have been found at low frequencies. This suggests that this population of pulsars may have particularly steep spectral indices. Continued multi-frequency study of pulsars discovered in unassociated LAT sources will deny or confirm this hypothesis.

%% file: Conclusions.tex
\section{Conclusion}
\label{sec:conclusions}  

We have performed repeated observations of 289 unassociated sources from the First \emph{Fermi} LAT catalogue of
$\gamma$-ray sources, resulting in the discovery of PSR
J1745$+$1017.

PSR J1745$+$1017 has a 2.65-ms spin period and orbits a very low-mass
companion with an orbital period of 17.5 hours. The low mass of the
companion suggests that PSR J1745$+$1017 is a Black Widow system,
however a two-year timing campaign has thus far found no evidence of
eclipses, dispersion measure variations across orbital phase or
Shapiro delay. By performing weighted-probability analysis of LAT photons
in the region of interest around PSR J1745$+$1017, we also detected
$\gamma$-ray pulsations from the source. This gives a clear indication
that PSR J1745$+$1017 is responsible for the $\gamma$-ray emission
seen from 1FGL~J1745.5$+$1018 (2FGL~J1745.6$+$1015).

The observed sample contains three radio pulsars, newly discovered
through similar searches performed by other observatories, and four
$\gamma$-ray selected, radio-quiet pulsars, found through blind
searches of \emph{Fermi} LAT data.

No radio detection was made for the four $\gamma$-ray selected pulsars
contained in our sample. The upper limits on their radio flux
densities at 1.36 GHz were calculated. These limits can be found in
Table \ref{tab:missed-gamma}.

Of the three newly discovered radio pulsars coincident with LAT
sources searched in this survey, only PSR J2030$+$3641 was detected
using the known timing solution. PSRs J0307$+$7443 and J1828$+$0625
were suspected to have radio flux densities at 1.36 GHz lower than the
limiting flux density of our observations.

%% file: PointingsAll2.tex
\begin{appendices}
   \section{Sample pointing information}
\pagestyle{empty}
\onecolumn
\centering
 \begin{scriptsize}
\begin{landscape}
\begin{longtable}{cllcccccccccccc}
\caption{Observed sources as ranked by the 2FGL logarithmic
  likelihood, 2FGL $R_{\rm{s}}$, found through application of
  Gaussian-mixture modeling \citep{Lee2012} to the 2FGL
  catalogue. $R_{95}$ values are the major-axes of the 95\% error
  ellipses for each source as found in the 1FGL and 2FGL
  catalogues. Limiting flux densities are calculated by
  the method outlined in Section \ref{sec:sensitivity}, with a 10\%
  duty-cycle imposed. Both Galactic coordinates, $l$ and $b$, and
  equatorial coordinates, R.A. and Decl., show the true pointing position of each observation.\label{tab:alldata}}\\ \hline
\hline
Index & 1FGL name & 2FGL name & l & b & R.A.  & Decl. & Offset from &
  1FGL $R_{95}$ & 2FGL $R_{95}$ & Observations & Limiting flux & 1FGL
  $R_{\rm{s}}$ & 2FGL $R_{\rm{s}}$\\ 
& & & ($^{\circ}$) & ($^{\circ}$) & J2000 & J2000 & 2FGL ($^{\circ}$)
  & ($^{\circ}$) & ($^{\circ}$) & (mins) & density (mJy) & & \\ 
\hline
\hline
1  &  J1906.6+0716c  &  J1906.5+0720  & 41.13 & -0.03 & 19:06:36             & $+$07:16:41            & 0.06 &  0.05  &  0.06  &  32  &  0.10  &  11.9  &  18.2 \\
  &   &   & 41.13 & -0.03 &  19:06:36  &  $+$07:16:41  & 0.06 &   &   &  10  &  0.18  &   &  \\
  &   &   & 41.13 & -0.03 &  19:06:36  &  $+$07:16:41  & 0.06 &   &   &  60  &  0.07  &   &  \\
  &   &   & 41.13 & -0.03 &  19:06:36  &  $+$07:16:41  & 0.06 &   &   &  60  &  0.07  &   &  \\
\hline
2  &  J1819.4-1518c  &  J1819.3-1523  & 15.70 & -0.05 & 18:19:25             & $-$15:18:26            & 0.09 &  0.12  &  0.12  &  36  &  0.09  &  7.27  &  15.5 \\
\hline
\end{longtable}
\end{landscape}
\end{scriptsize}
\end{appendices}

%% file: master_page.bbl
\begin{thebibliography}{}

\bibitem[\protect\citeauthoryear{Abdo et~al.}{Abdo et~al.}{2010a}]{Abdo2010}
Abdo A.~A. et~al., 2010a, The Astrophysical Journal Supplement Series, 188, 405

\bibitem[\protect\citeauthoryear{Abdo et~al.}{Abdo et~al.}{2010b}]{Abdo2010a}
Abdo A.~A. et~al., 2010b, The Astrophysical Journal Supplement Series, 187, 460

\bibitem[\protect\citeauthoryear{Atwood et~al.}{Atwood
  et~al.}{2009}]{Atwood2009}
Atwood W.~B. et~al., 2009, The Astrophysical Journal, 697/P, 1071

\bibitem[\protect\citeauthoryear{Burgay et~al.}{Burgay
  et~al.}{2006}]{Burgay2006a}
Burgay M. et~al., 2006, Monthly Notices of the Royal Astronomical Society, 368,
  283

\bibitem[\protect\citeauthoryear{Camilo et~al.}{Camilo
  et~al.}{2012}]{Camilo2012}
Camilo F. et~al., 2012, The Astrophysical Journal, 746, 39

\bibitem[\protect\citeauthoryear{Camilo, Thorsett, \& Kulkarni}{Camilo
  et~al.}{1994}]{Camilo1994}
Camilo F., Thorsett S.~E.,  Kulkarni S.~R., 1994, The Astrophysical Journal,
  421, L15

\bibitem[\protect\citeauthoryear{Champion, McLaughlin, \& Lorimer}{Champion
  et~al.}{2005}]{Champion2005}
Champion D.~J., McLaughlin M.~A.,  Lorimer D.~R., 2005, Monthly Notices of the
  Royal Astronomical Society, 364, 1011

\bibitem[\protect\citeauthoryear{Cognard et~al.}{Cognard
  et~al.}{2011}]{Cognard2011}
Cognard I. et~al., 2011, The Astrophysical Journal, 732, 47

\bibitem[\protect\citeauthoryear{Cordes \& Lazio}{Cordes \&
  Lazio}{2002}]{Cordes2002}
Cordes J.~M.,  Lazio T.~J.~W., 2002, arXiv:astro-ph/0207156v3, 21

\bibitem[\protect\citeauthoryear{Daugherty \& Harding}{Daugherty \&
  Harding}{1986}]{Daugherty1986}
Daugherty J.~K.,  Harding A.~K., 1986, The Astrophysical Journal, 309, 362

\bibitem[\protect\citeauthoryear{de~Jager \& B\"{u}sching}{de~Jager \&
  B\"{u}sching}{2010}]{deJager2010}
de~Jager O.~C.,  B\"{u}sching I., 2010, Astronomy and Astrophysics, 517, L9

\bibitem[\protect\citeauthoryear{Eatough et~al.}{Eatough
  et~al.}{2010}]{Eatough2010}
Eatough R.~P., Molkenthin N., Kramer M., Noutsos A., Keith M.~J., Stappers
  B.~W.,  Lyne A.~G., 2010, Monthly Notices of the Royal Astronomical Society,
  407, 2443

\bibitem[\protect\citeauthoryear{Edwards et~al.}{Edwards
  et~al.}{2001}]{Edwards2001}
Edwards R., Bailes M., van Straten W.,  Britton M., 2001, Monthly Notices of
  the Royal Astronomical Society, 326, 358

\bibitem[\protect\citeauthoryear{Faucher‐-Gigu\`{e}re \&
  Kaspi}{Faucher‐-Gigu\`{e}re \& Kaspi}{2006}]{FaucherGiguere2006}
Faucher‐-Gigu\`{e}re C.,  Kaspi V.~M., 2006, The Astrophysical Journal, 643,
  332

\bibitem[\protect\citeauthoryear{Foster \& Backer}{Foster \&
  Backer}{1990}]{Foster1990}
Foster R.~S.,  Backer D.~C., 1990, The Astrophysical Journal, 361, 300

\bibitem[\protect\citeauthoryear{Fruchter, Stinebring, \& Taylor}{Fruchter
  et~al.}{1988}]{Fruchter1988}
Fruchter A.~S., Stinebring D.~R.,  Taylor J.~H., 1988, Nature, 333, 237

\bibitem[\protect\citeauthoryear{Grindlay}{Grindlay}{1972}]{Grindlay1972}
Grindlay J.~E., 1972, The Astrophysical Journal, 174, L9

\bibitem[\protect\citeauthoryear{Guillemot et~al.}{Guillemot
  et~al.}{2012}]{Guillemot2012}
Guillemot L. et~al., 2012, The Astrophysical Journal, 744, 33

\bibitem[\protect\citeauthoryear{Hobbs, Edwards, \& Manchester}{Hobbs
  et~al.}{2006}]{Hobbs2006}
Hobbs G.~B., Edwards R.~T.,  Manchester R.~N., 2006, Monthly Notices of the
  Royal Astronomical Society, 369, 655

\bibitem[\protect\citeauthoryear{Jacoby et~al.}{Jacoby
  et~al.}{2009}]{Jacoby2009}
Jacoby B.~A., Bailes M., Ord S.~M., Edwards R.~T.,  Kulkarni S.~R., 2009, The
  Astrophysical Journal, 699, 2009

\bibitem[\protect\citeauthoryear{Kanbach et~al.}{Kanbach
  et~al.}{1989}]{Kanbach1989}
Kanbach G. et~al., 1989, Space Science Reviews, 49, 69

\bibitem[\protect\citeauthoryear{Kaplan et~al.}{Kaplan
  et~al.}{2012}]{Kaplan2012}
Kaplan D.~L. et~al., 2012, arXiv:1205.3699, 10

\bibitem[\protect\citeauthoryear{Keith et~al.}{Keith et~al.}{2008}]{Keith2008}
Keith M.~J., Johnston S., Kramer M., Weltevrede P., Watters K.~P.,  Stappers
  B.~W., 2008, Monthly Notices of the Royal Astronomical Society, 389, 1881

\bibitem[\protect\citeauthoryear{Keith et~al.}{Keith et~al.}{2011}]{Keith2011}
Keith M.~J. et~al., 2011, Monthly Notices of the Royal Astronomical Society,
  414, 1292

\bibitem[\protect\citeauthoryear{Kerr}{Kerr}{2011}]{Kerr2011}
Kerr M., 2011, The Astrophysical Journal, 732, 38

\bibitem[\protect\citeauthoryear{Kouwenhoven \& Vo\^{u}te}{Kouwenhoven \&
  Vo\^{u}te}{2001}]{Kouwenhoven2001}
Kouwenhoven M.~L.~A.,  Vo\^{u}te J.~L.~L., 2001, Astronomy and Astrophysics,
  378, 700

\bibitem[\protect\citeauthoryear{Kramer et~al.}{Kramer
  et~al.}{1998}]{Kramer1998}
Kramer M., Xilouris K.~M., Lorimer D.~R., Doroshenko O., Jessner A.,
  Wielebinski R., Wolszczan A.,  Camilo F., 1998, The Astrophysical Journal,
  501, 270

\bibitem[\protect\citeauthoryear{Lange et~al.}{Lange et~al.}{2001}]{Lange2001}
Lange C., Camilo F., Wex N., Kramer M., Backer D., Lyne A.,  Doroshenko O.,
  2001, Monthly Notices of the Royal Astronomical Society, 326, 274

\bibitem[\protect\citeauthoryear{Lee et~al.}{Lee et~al.}{2012}]{Lee2012}
Lee K.~J., Guillemot L., Yue Y.~L., Kramer M.,  Champion D.~J., 2012, Monthly
  Notices of the Royal Astronomical Society, 424, 2832

\bibitem[\protect\citeauthoryear{Lorimer \& Kramer}{Lorimer \&
  Kramer}{2005}]{handbook}
Lorimer D.,  Kramer M., 2005.
\newblock Cambridge University Press

\bibitem[\protect\citeauthoryear{Lorimer et~al.}{Lorimer
  et~al.}{2006}]{Lorimer2006}
Lorimer D.~R. et~al., 2006, Monthly Notices of the Royal Astronomical Society,
  372, 777

\bibitem[\protect\citeauthoryear{Manchester et~al.}{Manchester
  et~al.}{2001}]{Manchester2001}
Manchester R. et~al., 2001, Monthly Notices of the Royal Astronomical Society,
  328, 17

\bibitem[\protect\citeauthoryear{Nolan et~al.}{Nolan et~al.}{2012}]{Nolan2012}
Nolan P.~L. et~al., 2012, The Astrophysical Journal Supplement Series, 199, 31

\bibitem[\protect\citeauthoryear{Pletsch et~al.}{Pletsch
  et~al.}{2012}]{Pletsch2012}
Pletsch H.~J. et~al., 2012, The Astrophysical Journal, 744, 105

\bibitem[\protect\citeauthoryear{Ransom}{Ransom}{2001}]{Ransom2001}
Ransom S.~M., 2001, Ph.D. thesis, Harvard University

\bibitem[\protect\citeauthoryear{Ransom, Eikenberry, \& Middleditch}{Ransom
  et~al.}{2002}]{Ransom2002}
Ransom S.~M., Eikenberry S.~S.,  Middleditch J., 2002, The Astronomical
  Journal, 124, 1788

\bibitem[\protect\citeauthoryear{Ransom et~al.}{Ransom
  et~al.}{2011}]{Ransom2011}
Ransom S.~M. et~al., 2011, The Astrophysical Journal, 727, L16

\bibitem[\protect\citeauthoryear{Ray et~al.}{Ray et~al.}{2012}]{Ray2012}
Ray P.~S. et~al.

\bibitem[\protect\citeauthoryear{Ray et~al.}{Ray et~al.}{2011}]{Ray2011a}
Ray P.~S. et~al., 2011, The Astrophysical Journal Supplement Series, 194, 17

\bibitem[\protect\citeauthoryear{Ray \& {Saz Parkinson}}{Ray \& {Saz
  Parkinson}}{2011}]{Ray2011b}
Ray P.~S.,  {Saz Parkinson} P.~M., 2011, in High-Energy Emission from Pulsars
  and their Systems, Astrophysics and Space Science Proceeding, p.~21

\bibitem[\protect\citeauthoryear{Romani \& Yadigaroglu}{Romani \&
  Yadigaroglu}{1995}]{Romani1995}
Romani R.~W.,  Yadigaroglu I.-A., 1995, The Astrophysical Journal, 438, 314

\bibitem[\protect\citeauthoryear{Ruderman \& Sutherland}{Ruderman \&
  Sutherland}{1975}]{Ruderman1975}
Ruderman M.~A.,  Sutherland P.~G., 1975, The Astrophysical Journal, 196, 51

\bibitem[\protect\citeauthoryear{{Saz Parkinson}}{{Saz
  Parkinson}}{2011}]{Parkinson2011}
{Saz Parkinson} P.~M., 2011, in arXiv:1101.3096, p.~48

\bibitem[\protect\citeauthoryear{Shklovskii}{Shklovskii}{1970}]{Shklovskii1970}
Shklovskii I., 1970, Soviet Astronomy, 13, 562

\bibitem[\protect\citeauthoryear{Stappers et~al.}{Stappers
  et~al.}{1996}]{Stappers1996}
Stappers B.~W. et~al., 1996, The Astrophysical Journal, 465, L119

\bibitem[\protect\citeauthoryear{Thompson}{Thompson}{2008}]{Thompson2008a}
Thompson D.~J., 2008, Reports on Progress in Physics, 71, 116901

\bibitem[\protect\citeauthoryear{Vasseur et~al.}{Vasseur
  et~al.}{1970}]{Vasseur1970}
Vasseur J. et~al., 1970, Nature, 226, 534

\bibitem[\protect\citeauthoryear{Venter, Harding, \& Guillemot}{Venter
  et~al.}{2009}]{Venter2009}
Venter C., Harding A.~K.,  Guillemot L., 2009, The Astrophysical Journal, 707,
  800

\bibitem[\protect\citeauthoryear{Watters et~al.}{Watters
  et~al.}{2009}]{Watters2009}
Watters K.~P., Romani R.~W., Weltevrede P.,  Johnston S., 2009, The
  Astrophysical Journal, 695, 1289

\end{thebibliography}
